\documentclass[12pt]{iopart}
\input epsf 
\usepackage[dvips]{graphicx}

\begin{document}
\title{Choreographic Three Bodies on the Lemniscate}
\author{Toshiaki Fujiwara\dag, Hiroshi Fukuda\ddag\ and Hiroshi Ozaki\P}
\address{\dag\ Faculty of General Studies, Kitasato University, 
Kitasato 1-15-1, Sagamihara, Kanagawa 228-8555, Japan}%
\address{\ddag\ School of Administration and Informatics,
University of Shizuoka, 
52-1 Yada, Shizuoka 422-8526, Japan}
\address{\P\ Department of Physics, Tokai University,
1117 Kitakaname, Hiratsuka, Kanagawa 259-1292, Japan}

\eads{
	\mailto{\dag\ fujiwara@clas.kitasato-u.ac.jp},
	\mailto{\ddag\ fukuda@u-shizuoka-ken.ac.jp},
	\mailto{\P\ ozaki@keyaki.cc.u-tokai.ac.jp}
}

\begin{abstract}
We show that choreographic three bodies
$\{\mathbf{x}(t), \mathbf{x}(t+T/3), \mathbf{x}(t-T/3)\}$
of period $T$ 
on the lemniscate, 
$\mathbf{x}(t) =(\hat{\mathbf{x}}+\hat{\mathbf{y}}\mathrm{cn}(t))\mathrm{sn}(t)/(1+\mathrm{cn}^2(t))$
parameterized by 
the Jacobi's elliptic functions $\mathrm{sn}$ and $\mathrm{cn}$
with modulus $k^2 = (2+\sqrt{3})/4$,
conserve the center of mass 
and
the 
angular momentum, 
where $\hat{\mathbf{x}}$ and $\hat{\mathbf{y}}$ are 
the orthogonal unit vectors defining the plane of the motion.
They also conserve 
the
moment of inertia,
the
kinetic energy,
the
sum of square of the curvature, 
the product of distance and
the sum of square of distance between bodies.
We find that
they satisfy the equation of motion 
under the potential energy
$\sum_{i<j}(1/2 \ln r_{ij} -\sqrt{3}/24 r_{ij}^2)$
or
$\sum_{i<j}1/2 \ln r_{ij} -\sum_{i}\sqrt{3}/8 r_{i}^2$,
where $r_{ij}$ the distance between the body $i$ and $j$, and
$r_{i}$ the distance from the origin.
The first term of the potential energies 
is the Newton's gravity in two dimensions but 
the second term is the mutual repulsive force or 
a repulsive force from the origin, respectively. 
Then, geometric construction methods for the positions
of the choreographic three bodies are given.
\end{abstract}


%
%

\section{Introduction}
Choreographic motion of $N$ bodies is
a periodic motion
on a closed orbit,
$N$ bodies chase each other  on this orbit
with equal time-spacing.
Recently,
choreographic motions under the Newton's gravity
are found and paid attentions.
C. Moore\cite{moore} found a figure eight three-body
choreographic solution by numerical calculations.
A. Chenciner and R. Montgomery\cite{chenAndMont}
gave a rigorous proof of the existence of
the choreographic figure eight three-body solution.
At the same time,
C. Sim\'{o}\cite{simo1,simo2} found many remarkable
choreographic $N$-body solutions
by numerical calculations.
Although exact form of these choreographic $N$-body solutions
are still unknown,
Sim\'{o}'s figure eight choreographic three-body solution is very similar
to an affine transformed lemniscate\cite{simo2}.

Concerning the relation between the lemniscate
and the Newton's equation of motion,
it is well known\cite{Spiegel} that a point particle
on the lemniscate $r^2=\cos(2\theta)$
with $\theta=1/2\sin^{-1}(2lt)$, $-1\leq 2lt \leq 1$
satisfies the equation of motion
under the central potential $U(r)=-l^2/(2r^6)$.
Here $r$, $\theta$, $l$, and $t$ represent
the radius, the azimuthal angle, the angular momentum of the particle
and the time, respectively.
This motion is not periodic.
This particle starts from the origin
at $t=-1/(2l)$, travels right leaf of the lemniscate,
and finally collides with the origin at $t=1/(2l)$.
Two-body problem on the lemniscate is derived from the one body problem.
Two particles of equal masses
start from the origin at $t=-1/(2l)$ to opposite directions,
each particle travels the left leaf or right leaf of the half size lemniscate,
and collides with each other at the origin at $t=1/(2l)$.
No analytic solution is known about more than three bodies on the lemniscate.

Being stimulated by the
Moore, Chenciner, Montgomery and Sim\'{o}'s
remarkable study,
we investigated physical and geometrical properties
of the choreographic motion of
three bodies on the lemniscate.
We describe the results in section 2
and proof is given in section \ref{sec:proof}.
In section 4, we discuss
the relation between the choreographic three bodies on the lemniscate
and the rectangular hyperbola, and give a geometrical method to
determine the positions of the choreographic three bodies.
Section 5 is the summary.
%

\section{Choreographic three bodies on the lemniscate}
Let us parameterize 
by using the Jacobi's elliptic functions $\mathrm{sn}$ and $\mathrm{cn}$
the lemniscate
$\mathbf{x}(t)=x(t)\hat{\mathbf{x}} + y(t)\hat{\mathbf{y}}$
which satisfies $(x^2+y^2)^2=x^2-y^2$
as follows,
\begin{equation}
\label{defofxy}
\left\{
	\eqalign{
	x(t) &= \frac{\mathrm{sn}(t)}{1+\mathrm{cn}^2(t)},\\
	y(t) &= \frac{\mathrm{sn}(t)\mathrm{cn}(t)}{1+\mathrm{cn}^2(t)},
	}
\right.
\end{equation}
where $\hat{\mathbf{x}}$ and $\hat{\mathbf{y}}$ are 
the two basic orthogonal unit vectors 
defining the plane of the motion.
This is a smooth periodic motion on the lemniscate with period 
\begin{equation}
T=4K,
\end{equation}
where $K$ is the complete elliptic integral of the first kind.
Positions of the choreographic three bodies are
\begin{equation}
\label{threebody}
\{\mathbf{x}_{1}(t), \mathbf{x}_{2}(t), \mathbf{x}_{3}(t)\}
=\{\mathbf{x}(t), \mathbf{x}(t+4K/3), \mathbf{x}(t-4K/3)\}.
\end{equation}
In the following, we use notations 
$\mathbf{v}(t)=\rmd\mathbf{x}(t)/\rmd t$ and
$\mathbf{a}(t)=\rmd^{2}\mathbf{x}(t)/\rmd t^{2}$.

Straight calculation shows a relation between the curvature
\[
\rho^{-1}(t)=\frac{|\mathbf{v}(t)\times\mathbf{a}(t)|}{|\mathbf{v}(t)|^{3}}
\]
and the distance from the origin,
\begin{equation}
\label{eq:curvature}
	\rho^{-2}(t) = 9 \mathbf{x}^{2}(t),
\end{equation}
which is, of course, parameterization invariant relation. 
And also, we get a relation 
between the velocity and the distance from the origin,
\begin{equation}
\label{eq:harmonic}
	\mathbf{v}^{2}(t)
		+(k^2-\frac{1}{2})\mathbf{x}^2(t)=\frac{1}{2},
\end{equation}
for arbitrary modulus $k$.

\begin{figure}
\begin{center}
\epsfbox{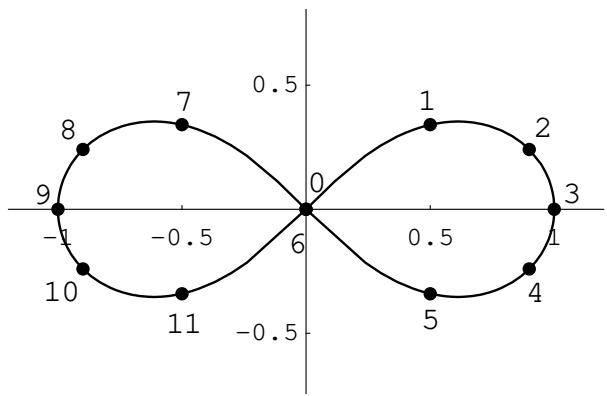}
\end{center}
\caption{\label{fig:eight}
The lemniscate and the positions of 
$\mathbf{x}(t)$
with modulus $k^2=(2+\sqrt{3})/4$
at $t=jK/3$ for $j=0,1,2,...,11$.
Full circles \fullcircle labeled by $j$ represent the positions of $\mathbf{x}(t)$. 
}
\end{figure}

Conservation of the center of mass
\begin{equation}
\label{centerofmass}
\mathbf{x}(t)+\mathbf{x}(t+4K/3)+\mathbf{x}(t-4K/3)=\mathbf{0}
\end{equation}
is satisfied if and only if the value of the modulus is 
\begin{equation}
\label{valueOfk2}
k^2=\frac{2+\sqrt{3}}{4},
\end{equation}
as shown in section \ref{sec:proof}.
Parameterization in \eref{defofxy}, \eref{threebody} and 
\eref{valueOfk2} defines the motion of
the choreographic three bodies on the lemniscate.
In \fref{fig:eight}, the lemniscate and 
the positions $\mathbf{x}(t)$ at $t=jK/3$ for $j=0,1,2,...,11$ are shown.
The order of magnitude of the label $j$
represents the direction of the motion and 
the points having the same label in modulo 4 are
the positions of the choreographic three bodies \eref{threebody} on the lemniscate at
$t=0, K/3, 2K/3, K, ...$.

We find that this motion conserves the moment of inertia and angular momentum
\begin{eqnarray}
 \sum \mathbf{x}_{i}^2 = \sqrt{3}, \label{eq:inertia}\\
 \sum \mathbf{x}_{i}\times\mathbf{v}_{i}=0, \label{eq:angular}
\end{eqnarray}
as shown in section \ref{sec:proof}.
Conservation of the moment of inertia and relations
\eref{eq:curvature}, \eref{eq:harmonic}
yield the conservation of the sum of square of the curvature and the kinetic energy,
\begin{eqnarray}
\sum \rho^{-2}_{i} = 9\sqrt{3},\label{eq:consOfCurvature}\\
\sum \mathbf{v}_{i}^2 = \frac{3}{4}. \label{eq:kinetic}
\end{eqnarray}
Simple algebra shows that \eref{centerofmass} implies  
$\sum_{i < j} (\mathbf{x}_{i}-\mathbf{x}_{j})^2 = 3 \sum_{i} \mathbf{x}_{i}^2$.
Thus we get another conservation,
\begin{equation}
\label{sofsdist}
\sum_{i < j} (\mathbf{x}_{i}-\mathbf{x}_{j})^2 = 3 \sqrt{3}.
\end{equation} 

The affine transformed motion 
$x(t)\hat{\mathbf{x}}+k^2 y(t)\hat{\mathbf{y}}$ 
is numerically close to the
Sim\'{o}'s figure eight motion.
It is difficult to distinguish the two orbits if they are plotted 
on the usual computer display.
This similarity and the above conservation laws suggest that 
the choreographic three bodies on the lemniscate \eref{defofxy}, \eref{threebody}
and \eref{valueOfk2}
may satisfy
an equation of motion under a interaction potential energy similar to that 
of the Newton's gravity.
We find that it surely satisfies an equation of motion
under the Newton's gravity in two dimensions with extra repulsive force,
\begin{equation}
\label{eqofmotion}
\eqalign{
	\frac{\rmd^2}{\rmd t^2}\mathbf{x}(t)
		&=\mathbf{F}_{\mathrm{Newton}}+\mathbf{F}_{\mathrm{repulsive}},\\
	\mathbf{F}_{\mathrm{Newton}}&=\frac{1}{2}
		\left\{
			\frac{\mathbf{x}(t+4K/3)-\mathbf{x}(t)}
				{\left(\mathbf{x}(t+4K/3)-\mathbf{x}(t)\right)^2}
			+
			\frac{\mathbf{x}(t-4K/3)-\mathbf{x}(t)}
				{\left(\mathbf{x}(t-4K/3)-\mathbf{x}(t)\right)^2}
	\right\},
}
\end{equation}
as shown in section \ref{sec:proof}.
The repulsive force $\mathbf{F}_{\mathrm{repulsive}}$ can be expressed in two ways
\begin{eqnarray}
\mathbf{F}_{\mathrm{repulsive1}}
	&=&\frac{\sqrt{3}}{4}\mathbf{x}(t),\label{eq:force1} \mbox{ or}\\
\mathbf{F}_{\mathrm{repulsive2}}
	&=&-\frac{\sqrt{3}}{12}
	\left\{
		\left(\mathbf{x}(t+\frac{4K}{3})-\mathbf{x}(t)\right)
		+\left(\mathbf{x}(t-\frac{4K}{3})-\mathbf{x}(t)\right)
	\right\} \label{eq:force2}
	.
\end{eqnarray}
Due to the conservation of the center of mass, the two repulsive force
is equal, $\mathbf{F}_{\mathrm{repulsive1}}=\mathbf{F}_{\mathrm{repulsive2}}$,
on the orbit.
The corresponding potential energy for the equation of motion with 
$\mathbf{F}_{\mathrm{repulsive1}}$ is 
\begin{equation}
\label{potential1}
 U=
 \sum_{i < j} \frac{1}{2}\ln r_{ij} 
 - \sum_{i}\frac{\sqrt{3}}{8}\mathbf{x}_{i}^2,
\end{equation}
and with $\mathbf{F}_{\mathrm{repulsive2}}$ is
\begin{equation}
V=
\sum_{i < j} 
\label{potential2}
	\left\{
		\frac{1}{2}\ln r_{ij} - \frac{\sqrt{3}}{24} r_{ij}^2
	\right\},
\end{equation}
where $r_{ij}$ is the 
distance
between the point $i$ and $j$, i.e., 
$r_{ij} = \sqrt{(\mathbf{x}_{i}-\mathbf{x}_{j})^2}$.

With the conservation of the kinetic energy and
the moment of inertia, the conservation of total energy
with potential energy $U$, in \eref{potential1},  implies the conservation of the
product of $r_{ij}^2$. Evaluating the value at $t=0$, we get
\begin{equation}
\label{pofdist}
r_{12}^2(t) r_{23}^2(t) r_{31}^2(t)= \frac{3\sqrt{3}}{2}.
\end{equation}


\section{Proof}
\label{sec:proof}
\subsection{Conservation of the center of mass}
Since $\mathbf{x}(K)=\hat{\mathbf{x}}$, 
the $\hat{\mathbf{x}}$-component of the other two points 
must be 
$x(K\pm 4K/3)=-1/2$. 
This equation gives,
\begin{equation}
\label{valueOfSn}
\mathrm{sn}(K/3)=\sqrt{3}-1.
\end{equation}
Evaluating $\mathrm{sn}(10K/3)$ in two ways,
%
\[
 \mathrm{sn}(10K/3)
   =\mathrm{sn}(3K+K/3)=-\mathrm{cn}(K/3)/\mathrm{dn}(K/3)
\]
and
\[
 \mathrm{sn}(10K/3)
   =\mathrm{sn}(4K-2K/3)
   =-\frac{2\mathrm{sn}(K/3)\mathrm{cn}(K/3)\mathrm{dn}(K/3)}
     {(1-k^2\mathrm{sn}^{4}(K/3))},
\]
we get
\begin{equation}
\label{kandsn}
k^2
=\frac{1-2\mathrm{sn}(K/3)}
	{\mathrm{sn}^4(K/3)-2\mathrm{sn}^3(K/3)}.
\end{equation}
Substituting the value of \eref{valueOfSn} 
into \eref{kandsn}
gives
the value of $k^2$ in \eref{valueOfk2}.
Inversely, the value of the modulus in
\eref{valueOfk2} gives the value of $\mathrm{sn}(K/3)$ 
in \eref{valueOfSn}. 
The value of $\mathrm{sn}(t), \mathrm{cn}(t), \mathrm{dn}(t)$ at 
some $t$ are shown in table~\ref{table:values}.
\begin{table}
\caption{Values of $\mathrm{sn}(t), \mathrm{cn}(t), \mathrm{dn}(t)$.}\label{table:values}
\begin{indented}
\item[]\begin{tabular}{@{}cccc}
\br
$t$	& $\mathrm{sn}(t)$	&$\mathrm{cn}(t)$	&$\mathrm{dn}(t)$\\
\mr
$K/3$&$\sqrt{3}-1$&$3^{1/4}(\sqrt{3}-1)/\sqrt{2}$&$1/\sqrt{2}$\\
$2K/3$&$3^{1/4}(\sqrt{3}-1)$&$2-\sqrt{3}$&$(\sqrt{3}-1)/2$\\
$4K/3$&$3^{1/4}(\sqrt{3}-1)$&$-2+\sqrt{3}$&$(\sqrt{3}-1)/2$\\
$5K/3$&$\sqrt{3}-1$&$-3^{1/4}(\sqrt{3}-1)/\sqrt{2}$&$1/\sqrt{2}$\\
\br
\end{tabular}
\end{indented}
\end{table}
%
To complete the proof of \eref{centerofmass}, it is convenient to 
use the complex variable 
\[
x^{(\pm)}(t)=x(t)\pm \rmi y(t) = \frac{\mathrm{sn}(t)}{(1\mp \rmi\ \mathrm{cn}(t))},
\]
and consider the complex $t$-plane.
We take the fundamental cell for $t$-plane with
\begin{equation}
 \label{eq:cell}
 -2K \leq \Re t < 2K, 
 -2K^{\prime} \leq \Im t < 2K^{\prime},
\end{equation}
where $K^{\prime}$ is the complementary elliptic integral of the first kind.
Since $x^{(\pm)}(t)$ has 4 simple zeros in the fundamental cell at
$t=-2K-2\rmi K^{\prime}, -2\rmi K^{\prime}, -2K$ and $0$, degree of the elliptic
functions $x^{(\pm)}(t)$ is 4.
Thus they should have 4 poles in the fundamental cell.
Note that,
\[
x^{(+)}(u+\rmi K^{\prime})=\frac{1}{k \mathrm{sn}(u) - \mathrm{dn}(u)}.
\]
From the value of $k^2$ in \eref{valueOfk2} and the table~\ref{table:values},
at $u=K/3$ and $5K/3$, $k\mathrm{sn}(u)=\mathrm{dn}(u)$. 
So, $x^{(+)}(t)$ has poles at $t=K/3+\rmi K^{\prime}$ and $5K/3+\rmi K^{\prime}$. 
Since $x^{(+)}(-t)=-x^{(+)}(t)$, it also has poles at 
$t=-(K/3+\rmi K^{\prime})$ and $-(5K/3+\rmi K^{\prime})$. 
We find 4 poles for degree 4
elliptic function $x^{(+)}(t)$,  then all these poles are simple poles.
Table~\ref{table:resOfxp} shows the poles and residues for 
$x^{(+)}(t)=\mathrm{sn}(t)/(1-\rmi\ \mathrm{cn}(t))$. 
Here we used the following notation,
\begin{equation}
\alpha_1 = -K+\rmi K^{\prime}, 
\alpha_2 = K/3+\rmi K^{\prime},
\alpha_3 = 5K/3+\rmi K^{\prime}.
\end{equation}
\begin{table}
\caption{Poles and residues of $\mathrm{sn}(t)/(1-\rmi\ \mathrm{cn}(t))$ 
and $1/(1-\rmi\ \mathrm{cn}(t))$.}\label{table:resOfxp}
\begin{indented}
\item[]\begin{tabular}{@{}ccccc}
\br
function	
		&$t=\alpha_2$	&$t=\alpha_3$
		&$t=-\alpha_2$	&$t=-\alpha_3$\\
\mr
$\mathrm{sn}(t)/(1-\rmi\ \mathrm{cn}(t))$	
		&$\sqrt{2}/3^{1/4}$  &-$\sqrt{2}/3^{1/4}$
		&$\sqrt{2}/3^{1/4}$  &-$\sqrt{2}/3^{1/4}$  \\
$1/(1-\rmi\ \mathrm{cn}(t))$
		&$1/3^{1/4}$  &-$1/3^{1/4}$
		&$-1/3^{1/4}$  &$1/3^{1/4}$  \\

\br
\end{tabular}
\end{indented}
\end{table}
It is clear from the table~\ref{table:resOfxp} that all these poles are cancelled in 
$x^{(+)}(t)+x^{(+)}(t+4K/3)+x^{(+)}(t-4K/3)$.
Therefore, it is constant. Evaluating this constant at $t=0$,
we get 
\begin{equation}
x^{(+)}(t)+x^{(+)}(t+4K/3)+x^{(+)}(t-4K/3)=0,
\end{equation}
which is equivalent to \eref{centerofmass}.
%
\subsection{Conservation of the moment of inertia and the angular momentum}
Let us consider a function 
\begin{equation}
 \label{func:ang}
 \frac{1}{1-\rmi\ \mathrm{cn}(t)}.
\end{equation}
Since
$1/(1-\rmi\ \mathrm{cn}(u+\rmi K^{\prime}))
=k \mathrm{sn}(u)/(k \mathrm{sn}(u)-\mathrm{dn}(u))
$,
the function \eref{func:ang}
has simple poles at $t=\pm\alpha_2, \pm\alpha_3$
in the above cell 
\eref{eq:cell}.
In table~\ref{table:resOfxp}, we list the poles and residues of this function.
From this table and the value at $t=0$, it is clear that 
the function \eref{func:ang}
satisfies
\begin{equation}
\label{eq:intj}
\frac{1}{1-\rmi\ \mathrm{cn}(t)}
+\frac{1}{1-\rmi\ \mathrm{cn}(t+4K/3)}
+\frac{1}{1-\rmi\ \mathrm{cn}(t-4K/3)}
= \frac{3+\sqrt{3}}{2}.
\end{equation}

Now, consider a function $j^{(+)}(t)=x^{(-)}(t)\rmd x^{(+)}(t)/\rmd t$.
Since $x^{(\pm)}=x\pm \rmi y$, 
\begin{equation}
\label{eq:j1}
j^{(+)}(t)=(x\frac{\rmd x}{\rmd t}+y\frac{\rmd y}{\rmd t})
+\rmi (x\frac{\rmd y}{\rmd t}-y\frac{\rmd x}{\rmd t})
.
\end{equation}
On the other hand, using $x^{(\pm)}(t)=\mathrm{sn}(t)/(1\mp \rmi\ \mathrm{cn}(t))$, we get
\begin{equation}
\label{eq:j2}
j^{(+)}(t)=\frac{\rmd}{\rmd t}\frac{1}{1-\rmi\ \mathrm{cn}(t)},
\end{equation} 
then, from equation~\eref{eq:intj}
\begin{equation}
\label{eq:j3}
j^{(+)}(t)+j^{(+)}(t+4K/3)+j^{(+)}(t-4K/3)=0.
\end{equation} 
Equations~\eref{eq:j1} and \eref{eq:j3} prove
the conservation of 
the moment of inertia and equation \eref{eq:angular}.
We get equation \eref{eq:inertia} from the value at $t=0$.

\subsection{Equation of motion}
Let us consider a function 
\[
\triangle x^{(-)}(t)=x^{(-)}(t+4K/3)-x^{(-)}(t).
\]
The function $x^{(-)}(t)$ has 4 simple poles 
at $\pm\alpha_{2}^{\ast}, \pm\alpha_{3}^{\ast}$,
the star being complex conjugation.
Then,  the function $\triangle x^{(-)}(t)$ has 6 simple poles at 
$\pm\alpha_{1}^{\ast}, \pm\alpha_{2}^{\ast}, \pm\alpha_{3}^{\ast}$.
So, the degree of this function is 6.

On $t=u+\rmi K^{\prime}$, the function $\triangle x^{(-)}(t)$ is
\[
\triangle x^{(-)}(u+\rmi K^{\prime})
=\frac{1}{k\mathrm{sn}(u+4K/3)+\mathrm{dn}(u+4K/3)}-\frac{1}{k\mathrm{sn}(u)+\mathrm{dn}(u)}.
\]
And from the table~\ref{table:values}, we see $\mathrm{sn}(5K/3)=\mathrm{sn}(K/3)$ 
and $\mathrm{dn}(5K/3)=\mathrm{dn}(K/3)$. 
Thus, at $u=K/3$, i.e., $t=K/3+\rmi K^{\prime}$, 
$\triangle x^{(-)}(t)$ has zero point.
To see the behaviour of the function around the zero point, let take $u=K/3+\triangle u$.
Then, using $\mathrm{sn}(2K-z)=\mathrm{sn}(z)$ and $\mathrm{dn}(2K-z)=\mathrm{dn}(z)$, we get
\[
	\triangle x^{(-)}(K/3+\triangle u+\rmi K^{\prime})
	=-\triangle x^{(-)}(K/3-\triangle u+\rmi K^{\prime}),
\]
i.e., $\triangle x^{(-)}(t)$ is an odd function around the zero point 
$t=\alpha_2=K/3+\rmi K^{\prime}$.

Series expansion of $x^{(-)}(u+4K/3+\rmi K^{\prime})$ 
and $x^{(-)}(u+\rmi K^{\prime})$ 
around $u=K/3$ yields
\begin{eqnarray}
\fl	x^{(-)}(u+4K/3+\rmi K^{\prime})\nonumber\\
\lo		=\frac{1}{\sqrt{2}}
		+\frac{1}{8}\sqrt{\frac{3}{2}}(u-K/3)^2\nonumber\\
		+\frac{3^{1/4}}{8\sqrt{2}}(u-K/3)^3
		+\frac{9}{128\sqrt{2}}(u-K/3)^4
		+\frac{3^{3/4}}{64\sqrt{2}}(u-K/3)^5
		+\cdots,
\end{eqnarray}
\begin{eqnarray}
\fl	x^{(-)}(u+\rmi K^{\prime})\nonumber\\
\lo	=\frac{1}{\sqrt{2}}
	+\frac{1}{8}\sqrt{\frac{3}{2}}(u-K/3)^2\nonumber\\
	-\frac{3^{1/4}}{8\sqrt{2}}(u-K/3)^3
	+\frac{9}{128\sqrt{2}}(u-K/3)^4
	-\frac{3^{3/4}}{64\sqrt{2}}(u-K/3)^5
	+\cdots.
\end{eqnarray}
Thus, we get
\begin{equation}
\triangle x^{(-)}(t)
=\frac{3^{1/4}}{4\sqrt{2}}(t-\alpha_2)^3
	+\frac{3^{3/4}}{32\sqrt{2}}(t-\alpha_2)^5+\cdots,
\end{equation}
and
\begin{equation}
\frac{1}{\triangle x^{(-)}(t)}
=\frac{4\sqrt{2}}{3^{1/4}}\frac{1}{(t-\alpha_2)^3}
	-\frac{3^{1/4}}{\sqrt{2}}\frac{1}{(t-\alpha_2)}+\cdots,
\end{equation}
i.e., $1/\triangle x^{(-)}(t)$ has triple pole at $t=\alpha_2$.
Since $x^{(-)}(-t)=-x^{(-)}(t)$, $\triangle x^{(-)}(t)$ also has triple zero at 
$t=-\alpha_3=-5K/3-\rmi K^{\prime}$,
as follows,
\[
	\triangle x^{(-)}(-\alpha_3+\triangle t)
	=\triangle x^{(-)}(\alpha_2-\triangle t)
\]
Thus $1/\triangle x^{(-)}(t)$ has triple pole at $t=-\alpha_3$
with principal part
\begin{equation}
\frac{1}{\triangle x^{(-)}(t)}
=-\frac{4\sqrt{2}}{3^{1/4}}\frac{1}{(t+\alpha_3)^3}
	+\frac{3^{1/4}}{\sqrt{2}}\frac{1}{(t+\alpha_3)}+\cdots.
\end{equation}
We found two triple poles for the elliptic function of degree 6.

In table~\ref{table:principalpart}, we list the principal part of 
$1/\triangle x^{(-)}(t)$, $-1/\triangle x^{(-)}(t-4K/3)$,
$\rmd^{2}x^{(+)}(t)/\rmd t^2$ and $x^{(+)}(t)$.
\begin{table}
\caption{Principal part around the pole with 
	$a=2\sqrt{2}/3^{1/4}$,
	$b=3^{1/4}/\sqrt{2}$.
}
\label{table:principalpart}
\begin{tabular}{@{}ccccc}
\br
function	
	&$t=\alpha_2$	&$t=\alpha_3$
	&$t=-\alpha_2$	&$t=-\alpha_3$\\
\mr
$[\triangle x^{(-)}(t)]^{-1}$
	&$\frac{2a}{(t-\alpha_2)^3}
	-\frac{b}{t-\alpha_2}$  
		&  
			&
				&$-\frac{2a}{(t+\alpha_3)^3}
					+\frac{b}{t+\alpha_3}$\\
$-[\triangle x^{(-)}(t-\frac{4K}{3}]^{-1}$
	&
		&$-\frac{2a}{(t-\alpha_3)^3}
			+\frac{b}{t-\alpha_3}$
			&$\frac{2a}{(t+\alpha_2)^3}
				-\frac{b}{t+\alpha_2}$
				&\\
$\rmd^{2}x^{(+)}(t)/\rmd t^2$
	&$\frac{a}{(t-\alpha_2)^3}$
		&$-\frac{a}{(t-\alpha_3)^3}$
			&$\frac{a}{(t+\alpha_2)^3}$
				&$-\frac{a}{(t+\alpha_3)^3}$\\
$x^{(+)}(t)$
	&$\frac{1}{b(t-\alpha_2)}$
		&$-\frac{1}{b(t-\alpha_3)}$
			&$\frac{1}{b(t+\alpha_2)}$
				&$-\frac{1}{b(t+\alpha_3)}$\\
\br
\end{tabular}
\end{table}
From this table, we see the following equation is satisfied,
\begin{equation}
\frac{\rmd^{2}}{\rmd t^2}x^{(+)}(t)=
	\frac{1}{2}
	\left\{
		\frac{1}{\triangle x^{(-)}(t)}-\frac{1}{\triangle x^{(-)}(t-\frac{4K}{3})}
	\right\}
	+\frac{\sqrt{3}}{4}x^{(+)}(t).
\end{equation}
This is the equation of motion in \eref{eqofmotion} with the repulsive force in
\eref{eq:force1}.

\section{Choreographic three bodies and the rectangular hyperbola}
In general three-body problem
on the plane, 
not restricted 
to this lemniscate case,
the conservation of the center of mass
and zero angular momentum has simple
geometrical meaning: three tangent lines from the three bodies
must meet at a point (infinity is allowed) at each instant.
That is, there exist scalars $\lambda_{i}(t)$ and a vector $\mathbf{c}(t)$,
such that,
\begin{equation}
\mathbf{c}=\mathbf{x}_{i} + \lambda_{i} \mathbf{v}_{i},
\mbox{ for } i = 1,2,3.
\end{equation} 
The solution 
$\lambda_{i}$ is,
\begin{equation}
\lambda_{i} = \frac{\{(\mathbf{x}_{j}-\mathbf{x}_{i}) \times \mathbf{v}_{j}\}\cdot\hat{\mathbf{z}}}
				{(\mathbf{v}_{i} \times \mathbf{v}_{j})\cdot\hat{\mathbf{z}}},
\mbox{ with } \forall j \neq i,
\end{equation}
and the vector $\mathbf{c}(t)$ is given by
\begin{equation}
\mathbf{c}
=-\frac{l_{i}\mathbf{v}_{j}-l_{j}\mathbf{v}_{i}}
              {(\mathbf{v}_{i}\times \mathbf{v}_{j})\cdot \hat{\mathbf{z}}},
\mbox{ with } (i,j)=(1,2),(2,3),(3,1),
\end{equation}
where $\hat{\mathbf{z}}=\hat{\mathbf{x}}\times\hat{\mathbf{y}}$ 
and
$l_{i} =(\mathbf{x}_{i} \times \mathbf{v}_{i})\cdot \hat{\mathbf{z}}$.

\begin{figure}
\begin{center}
\epsfbox{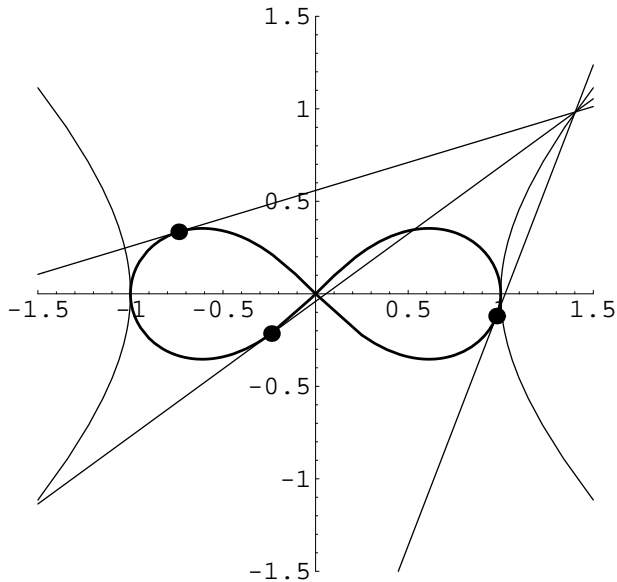}
\end{center}
\caption{\label{fig:hyperbola}
Snapshot at $t=-K/6$.
Three tangent lines from the choreographic
three bodies meet at a point
on the rectangular hyperbola.
Full circles \fullcircle
are positions of the three bodies.
}
\end{figure}

For the lemniscate, 
we find that
the orbit of $\mathbf{c}(t)$ is the rectangular hyperbola
as shown in \fref{fig:hyperbola}, 
\begin{equation}
(\mathbf{c}\cdot\hat{\mathbf{x}})^2 - (\mathbf{c}\cdot\hat{\mathbf{y}})^2 = 1
\end{equation}
by a direct algebraic calculation.


Let us introduce terms  ``forward''  and ``backward''  for the
direction of the choreographic motion on the lemniscate.
We call the motion ``forward'' if points pass through the origin upward
(from lower left to upper right or from lower right to upper left)
and call ``backward'' otherwise.
In other words, a ``forward'' motion is a clockwise motion on the right leaf
of the lemniscate or a anti-clockwise motion on the left leaf.

We observed the following two properties.
(i) For the ``forward'' motion of the choreographic three bodies,
the point $\mathbf{c}$ always moves upward on the rectangular hyperbola.
(ii) Four positions of the choreographic three bodies and
the cross point on the rectangular hyperbola,
$\mathbf{x}_{i}$, $i=1,2,3$ and $\mathbf{c}$,
are in the different quadrants each other
as shown in \fref{fig:hyperbola}.
Notice that when the point $\mathbf{c}$ jumps the leaf of the hyperbola
one of the three bodies passes through the origin.
And when the point $\mathbf{c}$ passes through the horizontal axis upward (downward)
one of the three bodies passes through
the horizontal axis at the same point in opposite direction, downward
(upward).

Then, we will have two questions as to 
how the choreographic three points can be determined geometrically.
The first question: 
How can we find the positions of the choreographic three bodies
for an arbitrarily given point $\mathbf{c}$ on the rectangular
hyperbola?
We can draw four (in general) tangent lines to the lemniscate
from the point $\mathbf{c}$.
Therefore we have four (in general) contact points instead of three.
How can we select the choreographic three points from among them?
We observed that the following two methods work.
(i)
Move the point $\mathbf{c}$ slightly upward.
Then, we will observe that three of the four
contact points move ``forward'',
and one ``backward''.
Of course the former is the positions of the choreographic three bodies.
(ii)
Choose three points which is not in the same quadrants
the given point $\mathbf{c}$ is in.

The second question: 
How can we find other two choreographic positions for an arbitrarily given
point $\mathbf{x}_{1}$ on the lemniscate?
The tangent line of the lemniscate which contact at the given point
$\mathbf{x}_{1}$ have two cross point,
$\mathbf{d}_1, \mathbf{d}_2$,
on the rectangular hyperbola.
How can we select one?
Again we have two method.
(i)
If we move the point $\mathbf{x}_{1}$ ``forward'',
one of $\mathbf{d}_i$ will move upward, and the other downward.
The former is the point $\mathbf{c}$ 
the point $\mathbf{x}_{1}$
corresponds to.
(ii)
Select a cross point $\mathbf{d}_{i}$ which is in the different quadrants
from the one the given point $\mathbf{x}_{1}$ is in.
Using the method described in the above paragraph,
we can find  positions
$\mathbf{x}_{2}$ and $\mathbf{x}_{3}$
for the other two choreographic bodies.

The above procedures give a geometrical method to find
a set of positions of the choreographic three bodies
on the lemniscate.

\section{Summary}
We have shown that the choreographic three bodies on the lemniscate
\eref{defofxy}, \eref{threebody} and \eref{valueOfk2}
satisfies the equation of motion under two different potential energies,
$U$ defined in \eref{potential1} and $V$ in \eref{potential2}.
It means that the choreographic three bodies on the lemniscate can be 
realized by the following two ways.
(i) Consider four particles,
the infinitely heavy particle 0 and three particles 1--3 of equal masses.
Take the Newton's gravity in two dimensions
\begin{equation}
 U_{ij}=\frac{1}{2}\ln{r_{ij}}
\end{equation}
as the interaction potential energy 
between particle $i$ and $j$ with $i,j=1,2,3$,
and 
the artificial repulsive potential energy
\begin{equation}
 U_{0i}=-\frac{\sqrt{3}}{8}\mathbf{x}_{i}^{2}
\end{equation}
as the interaction potential energy 
between the particle 0 and $i$ with $i=1,2,3$.
Then, put the infinitely heavy particle 0 at the origin and 
put the three particles 1--3 on the lemniscate according to equation \eref{threebody}.
(ii) Consider three particles 1--3 of equal masses.
Take the Newton's gravity in two dimensions accompanied 
with the artificial repulsive potential
\begin{equation}
 V_{ij}=\frac{1}{2}\ln{r_{ij}}-\frac{\sqrt{3}}{24}r_{ij}^{2}
\end{equation}
as the interaction potential energy 
between particle $i$ and $j$ with $i,j=1,2,3$.
Then, 
put the three particles 1--3 on the lemniscate according to equation \eref{threebody}.

As a consequence of the equation of motion, 
the choreographic three bodies on the lemniscate satisfy 
the kinematical conservation laws independent of the potential energies,
 i.e., 
the center of mass \eref{centerofmass} and 
the angular momentum \eref{eq:angular}.
Furthermore, they conserve a lot of kinematical and geometrical quantities, 
the kinetic energy \eref{eq:kinetic},
the moment of inertia \eref{eq:inertia},
the product of mutual distance \eref{pofdist},
the sum of square of mutual distance \eref{sofsdist}
and the sum of square of the curvature \eref{eq:consOfCurvature}.

We have also pointed out that
the tangent lines from the positions of the choreographic three bodies
on the lemniscate meet at a point on the rectangular hyperbola, then
have given the two geometrical methods
to determine the choreographic three points on the lemniscate.
We have given (i) how to determine the choreographic three bodies from
a point on the rectangular hyperbola, and
(ii) how to determine the choreographic three points from
the given first point on the lemniscate.



As we have already mentioned, we believed 
for some time
the choreographic three bodies on the lemniscate
was the Sim\'{o}'s figure eight motion because of the 
numerical similarity and a lot of conservation quantities of them.
We hope that 
we can discuss in the future the orbit of Moore, Chenciner, Montgomery
and Sim\'{o}'s figure eight motion
in our context.

\ack
We thank for Dr. Takehiko Fujishiro, 
Dr. Yumi Hirata,
Dr. Hideaki Hiro-oka, and
Dr. Hiroaki Arisue for their interests and discussions in our work.

\end{document}